\documentclass[a4paper]{jpconf}

\usepackage{amsmath}
\usepackage{graphicx}
\usepackage{bm}

\begin{document}
\title{Phase-Sensitive Flux-Flow resistivity in Unconventional Superconductors}

\author{Yoichi Higashi$^{a,b,d}$, Yuki Nagai$^{c,d,e}$, Masahiko Machida$^{c,d,e}$, and Nobuhiko Hayashi$^{b,d}$}

\address{$^{a}$Department of Mathematical Sciences, Osaka Prefecture University, 1-1 Gakuen-cho, Sakai 599-8531, Japan}
\address{$^{b}$Nanoscience and Nanotechnology Research Center (N2RC), Osaka Prefecture University, 1-2 Gakuen-cho, Sakai 599-8570, Japan}
\address{$^{c}$CCSE, Japan Atomic Energy Agency, 5-1-5 Kashiwanoha, Kashiwa, Chiba 277-8587, Japan}
\address{$^{d}$CREST(JST), 4-1-8 Honcho, Kawaguchi, Saitama 332-0012, Japan}
\address{$^{e}$TRIP, JST, 5 Sanban-cho, Chiyoda-ku, Tokyo 110-0075, Japan}

\ead{higashiyoichi@ms.osakafu-u.ac.jp}
\begin{abstract}
We theoretically investigate the magnetic-field-angle dependence of the flux-flow resistivity $\rho_{\rm f}$ in unconventional superconductors.
Two contributions to $\rho_{\rm f}$ are considered: 
one is the quasiparticle (QP) relaxation time $\tau(\bm{k}_{\rm F})$ and the other is $\omega_0(\bm{k}_{\rm F})$,
which is a counterpart to the interlevel spacing of the QP bound states in the quasiclassical approach.
Here, $\bm{k}_{\rm F}$ denotes the position on a Fermi surface.
Numerical calculations are conducted for a line-node $s$-wave and a $d$-wave pair potential with the same anisotropy of their amplitudes, but with a sign change only for a $d$-wave one.
We show that the field-angle dependence of $\rho_{\rm f}$ differs prominently between $s$-wave and $d$-wave pairs, reflecting the phase of the pair potentials.
We also discuss the case where $\tau$ is constant and compare it with the more general case where $\tau$ depends on $\bm{k}_{\rm F}$.
\end{abstract}
\section{Introduction}
　The Cooper pairing mechanism is reflected by the symmetry of the superconducting pair potential.
Therefore, the elucidation of the pair potential symmetry is of great importance for obtaining the clue to the Cooper pairing mechanism in unconventional superconductors \cite{Sigrist}.
There are two factors of the pair potential as a complex number: one is the amplitude and the other is the phase.
The magnetic-field-angle resolved thermal conductivity and specific heat measurements are powerful techniques which can detect the anisotropy of the pair potential amplitude \cite{Sakakibara,Hayashi}.
However, they cannot detect the phase of the pair potential.
It is crucial to probe its phase in order to discriminate unconventional sign reversed pair potential from conventional sign conserved one. 
To this end, we propose that the field-angle dependence of
the flux-flow resistivity $\rho_{\rm f}$ \cite{Yasuzuka} can be a phase-sensitive probe.

The flux-flow resistivity is
$\rho_{\rm f}(T) \propto \varGamma(\varepsilon=k_{\rm B}T)$ in moderately clean systems \cite{Kato}.
$\varepsilon$ is the energy of the quasiparticle (QP) bound state inside a vortex core and $T$ is the temperature.
In a previous paper \cite{Higashi},
we took into account only the QP scattering rate $\varGamma(\bm{k}_{\rm F})$ as a contribution to $\rho_{\rm f}$.
Here, the Fermi wave number 
$\bm{k}_{\rm F}$ denotes the position on a Fermi surface (FS).
Actually, there is another contribution to $\rho_{\rm f}$,
which is an energy scale $\omega_0(\bm{k}_{\rm F})$ related to the vortex bound state spectrum \cite{Kopnin}.
In this study,
we investigate the field-angle dependence of $\rho_{\rm f}$ 
taking into account the contribution of $\omega_0(\bm{k}_{\rm F})$ in addition to $\varGamma(\bm{k}_{\rm F})$.

\section{Formulation}
　We consider a single vortex at low magnetic fields.
We focus on the QP scattering due to the non-magnetic impurities distributed randomly inside a vortex core.
The QP scattering rate
$\varGamma$ is obtained by calculating the imaginary part of the impurity self energy ${\rm Im} \varSigma$,
which corresponds to the energy level width $\delta E$.
The QP scattering rate $\varGamma(\bm{k}_{\rm F})$ is related to $\delta E$ as
$\varGamma(\bm{k}_{\rm F}) \sim {\rm Im}\varSigma \sim \delta E \sim 1/\tau(\bm{k}_{\rm F})$.
In this paper, we set $\hbar=1$.
On the basis of the quasiclassical approximation \cite{Serene} and the Kramer-Pesch approximation \cite{Kramer}, 
the quasiclassical Green's functions and the impurity self energy
in the vicinity of a vortex core are obtained analytically \cite{Nagai}.
From the pole of the regular Green's function, 
the QP energy spectrum $E(y,\bm{k}_{\rm F})$ is obtained as $E(y,\bm{k}_{\rm F})=2y\vert d(\bm{k}_{\rm F}) \vert^2 \Delta^2_0/v_{\rm F \perp}$.
Here, 
$y$ is the impact parameter, 
$d(\bm{k}_{\rm F})$ is the anisotropy factor of the pair potential,
$\Delta_0$ is the maximum amplitude of the pair potential in the bulk,
and
$v_{\rm F \perp}$ is the FS average of $\vert \bm{v}_{\rm F \perp} (\bm{k}_{\rm F}) \vert$ defined in ref.\ \cite{Higashi}.
In the quasiclassical approximation, the QP spectrum is continuous with respect to the impact parameter $y$ \cite{Kopnin-text}.
The impact parameter $y(>0)$ is related to the angular momentum $l$ of the QPs running around a vortex core such that $-l=y\vert \bm{k}_{\rm F \perp} \vert$.
Here the negative angular momentum is due to the direction of the circular motion of the QPs around a vortex core.
The spectrum of the low-energy vortex bound states
is expressed \cite{Caloli,Volovik} as
$E(y,\bm{k}_{\rm F})=-\omega_0(\bm{k}_{\rm F})l=\omega_0(\bm{k}_{\rm F}) y \vert\bm{k}_{\rm F \perp} \vert$
in terms of a counterpart to the interlevel spacing of the QP bound states $\omega_0(\bm{k}_{\rm F})$.
Comparing the above two expressions of $E(y,\bm{k}_{\rm F})$,
$\omega_0(\bm{k}_{\rm F})$ is represented as 
\begin{equation}
\omega_0(\bm{k}_{\rm F})=\frac{   2 \vert d(\bm{k}_{\rm F}) \vert^2\Delta^2_0    }{   \vert \bm{k}_{\rm F \perp} \vert v_{\rm F \perp}  }.
\end{equation}

Then, $\rho_{\rm f}$ is given as \cite{Kopnin}
\begin{align}
 \rho_{\rm f}(T, \alpha_{\rm M}) & \propto \cfrac{1}{  \bigl\langle \omega_0(\bm{k}_{\rm F})\tau(\bm{k}_{\rm F}) \bigr\rangle_{\rm FS}  }
= \cfrac{1}{\left\langle \cfrac{2 \vert d(\bm{k}_{\rm F}) \vert^2}{  \pi \vert \bm{k}_{\rm F \perp} \vert \xi_0  } \cfrac{\varGamma_{\rm n}}{  \varGamma(\varepsilon=k_{\rm B}T, \bm{k}_{\rm F}, \alpha_{\rm M})  } \cfrac{\Delta_0}{  \varGamma_{\rm n}  } \right\rangle_{\rm FS}}\\
&=\frac{\pi}{2}k_{\rm F}\xi_0\frac{\varGamma_{\rm n}}{\Delta_0}\cfrac{1}{    \left\langle \cfrac{  \vert d(\bm{k}_{\rm F}) \vert^2  }{   \sqrt{  \cos^2\theta_k+\sin^2(\phi_k-\alpha_{\rm M})\sin^2\theta_k  }}\cfrac{  \varGamma_{\rm n}  }{  \varGamma(\varepsilon=k_{\rm B}T, \bm{k}_{\rm F}, \alpha_{\rm M})  }  \right\rangle_{\rm FS}},
\end{align}
where $\bm{k}_{\rm F \perp}$ is the component of $\bm{k}_{\rm F}$ projected onto the plane perpendicular to the field,
and
$\vert \bm{k}_{\rm F \perp} \vert=k_{\rm F}\sqrt{\cos^2\theta_k+\sin^2(\phi_k-\alpha_{\rm M})\sin^2\theta_k}$
with $\bm{k}_{\rm F}=k_{\rm F}(\cos\phi_k\sin\theta_k\hat{\bm{a}}+\sin\phi_k\sin\theta_k\hat{\bm{b}}+\cos\theta_k\hat{\bm{c}})$
in polar coordinates \cite{Higashi}.
$\hat{\bm{a}}$, $\hat{\bm{b}}$, $\hat{\bm{c}}$ are orthogonal unit vectors fixed to the crystal axes.
Here, we have assumed an isotropic spherical FS.
The angle $\alpha_{\rm M}$ indicates the direction of the applied magnetic field rotated in a plane perpendicular to the $c$ axis  \cite{Higashi},
and it is measured from the $\phi_k=0$ direction ($\hat{\bm{a}}$-axis direction).
$\varGamma_{\rm n}$ is the scattering rate in the normal state.
The coherence length is defined as $\xi_0=v_{\rm F \perp}/{\pi \Delta_0}$.
We use the weak-coupling BCS ratio of $\Delta_0 / k_{\rm B} T_{\rm c}=2/1.13$.
The brackets $\langle \cdots \rangle_{\mathop{\mathrm{FS}} }$ mean the FS integral
with respect to ${\bm k}_{\rm F}$ like,
$
\langle \cdots \rangle_{\mathop{\mathrm{FS}}}
\equiv (1/\nu_0)\int  dS_{\rm F} \cdots /\vert \bm{v}_{\rm F}(\bm{k}_{\rm F}) \vert 
=\int_0^{2\pi}d\phi_k \int_0^{\pi}d\theta_k\sin\theta_k \cdots /(4\pi)
$
with $dS_{\rm F}$ being a FS area element and $\nu_0=\int dS_{\rm F}/ \vert \bm{v}_{\rm F}(\bm{k}_{\rm F}) \vert$ being the density of states on the FS.
The expression for
$\varGamma(\varepsilon, \bm{k}_{\rm F}, \alpha_{\rm M})$ is given in ref.\ \cite{Higashi}.
%
\section{Results}
　We consider two Cooper pairing models on a spherical FS.
One is the line-node $s$-wave pair: $d(\bm{k}_{\rm F})=\vert \cos(2\phi_k) \vert \sin^2 \theta_k$.
The other is the $d_{x^2-y^2}$-wave pair: $d(\bm{k}_{\rm F})=\cos(2\phi_k)\sin^2 \theta_k$.
Both models have the same line nodes in the gap from the north pole of the Fermi sphere to the south one in the directions of $\phi_k=(1+2n)\pi/4$ with $n=0, 1, 2, 3$.
On the other hand, the anti-node directions correspond to $\phi_k=n\pi/4$.

\begin{figure}[h]
\begin{minipage}{16pc}
\includegraphics[width=20pc]{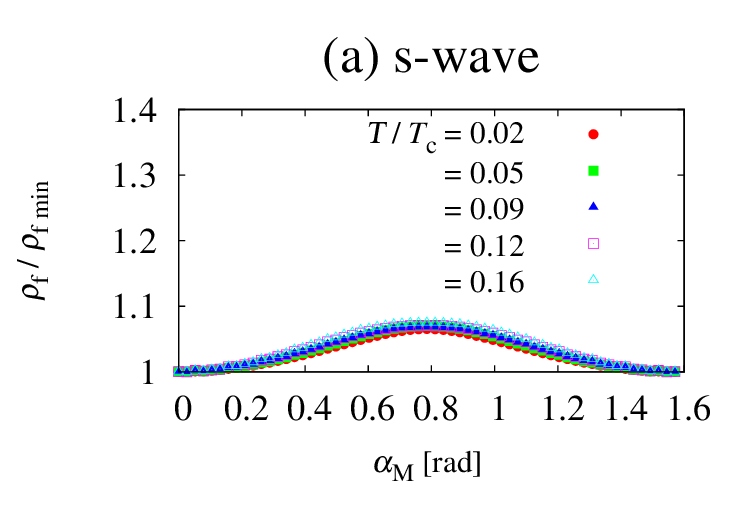}
\end{minipage}\hspace{2pc}%
\begin{minipage}{16pc}
\includegraphics[width=20pc]{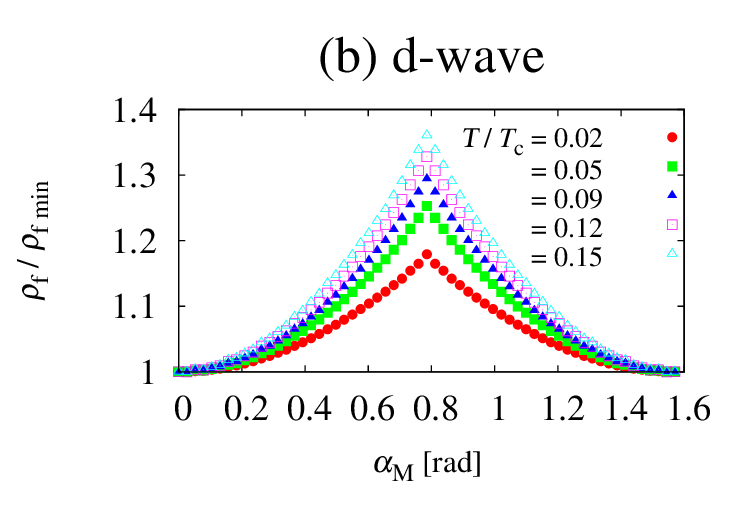}
\end{minipage} 
\caption{\label{fig1}The field-angle ($\alpha_{\rm M}$) dependence of the flux-flow resistivity $\rho_{\rm f}$ in the case of (a) the line-node $s$-wave pair and (b) the $d$-wave pair.
The data are plotted for each temperature $T$.
$T_{\rm c}$ is the superconducting critical temperature.
The vertical axis is normalized by minimum values $\rho_{\rm f~min}$ for each temperature.}
\end{figure}

In Fig.~\ref{fig1}, we show the field-angle ($\alpha_{\rm M}$) dependence of $\rho_{\rm f}$,
where each plot is normalized by its minimum $\rho_{\rm f~min}$ and therefore
the results do not depend on the parameter $k_{\rm F}  \xi_0$ or $\varGamma_{\rm n}/\Delta_0$.
As seen in Fig.~\ref{fig1}(a),
in the case of the line-node $s$-wave pair,
$\rho_{\rm f}$ exhibits its broad maximum when the magnetic field $\bm{H}$ is applied
parallel to the gap-node direction ($\alpha_{\rm M} \approx 0.8$).
It is noticed that  $\rho_{\rm f}$ has little temperature dependence.
On the other hand,
in the $d$-wave case [Fig.~\ref{fig1}(b)],
the sharp maximum appears when $\bm{H}$ is applied to the gap-node direction.
The ratio of the maximum value to the minimum one is much larger than
that of the $s$-wave case.
In addition,
the ratio increases with increasing the temperature $T$.

\begin{figure}[h]
\begin{center}
\includegraphics[width=18pc]{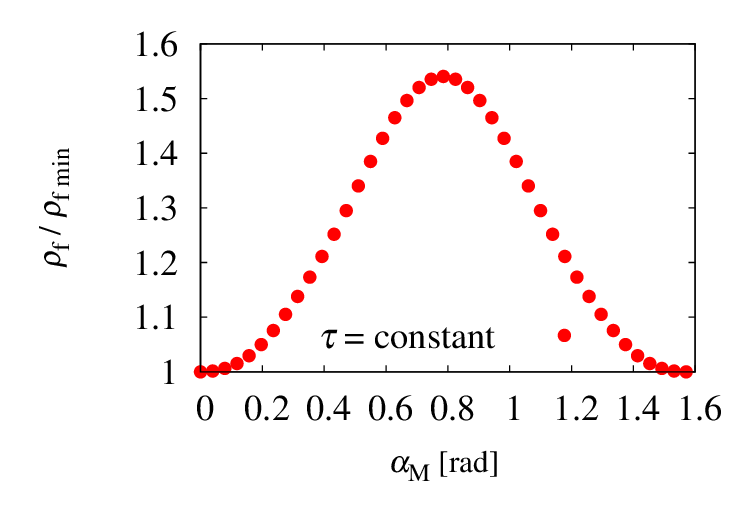}
\end{center}
\caption{\label{fig2}The field-angle ($\alpha_{\rm M}$) dependence of the flux-flow resistivity $\rho_{\rm f}$ in the case of $\tau=\mbox{const}$.
}
\end{figure}

The field-angle dependence of $\rho_{\rm f}$ in the case of the constant $\tau$ is displayed in Fig.~\ref{fig2}.
This corresponds to the case where we consider only the contribution of $\omega_0(\bm{k}_{\rm F})$ to $\rho_{\rm f}$
neglecting the $\bm{k}_{\rm F}$ dependence of the QP scattering rate $\varGamma$.
The maximum of $\rho_{\rm f}(\alpha_{\rm M})$ appears when the field is applied to the node direction, independent of
whether the pair potential is the line-node $s$-wave or the $d$-wave.
Such maximum direction $\alpha_{\rm M}$ in this case corresponds to the direction $\phi_k$ in the $k$ space
where $\omega_0(\bm{k}_{\rm F})$ is small.
The graph in Fig.~\ref{fig2} becomes flat if we consider an isotropic pairing $d(\bm{k}_{\rm F})=\mbox{const}$.

Combined with our previous results for $\varGamma(\alpha_{\rm M})$ \cite{Higashi}, 
the behavior of $\rho_{\rm f}(\alpha_{\rm M})$ can be understood as follows.
In the line-node $s$-wave case,
when the magnetic field is applied to the node direction,
$\varGamma(\alpha_{\rm M})$ takes the minimum \cite{Higashi}
while $\rho_{\rm f}(\alpha_{\rm M})$ due only to $\omega_0(\bm{k}_{\rm F})$ exhibits its maximum  [Fig.~\ref{fig2}].
As a result,
these two contributions to $\rho_{\rm f}$ cancel each other and only the small and broad maximum appears
[Fig.~\ref{fig1}(a)].
On the other hand,
in the $d$-wave case,
$\varGamma(\alpha_{\rm M})$ and $\rho_{\rm f}(\alpha_{\rm M})$ due to $\omega_0(\bm{k}_{\rm F})$ have
their maximum in the gap-node direction in common.
Therefore, the maximum of the field-angle dependence of $\rho_{\rm f}$ stands out and the sharp maximum appears [Fig.~\ref{fig1}(b)].

\section{Conclusion}
　We theoretically investigated the magnetic field-angle dependence of the flux-flow resistivity
for the line-node $s$-wave pair and the $d$-wave pair on a spherical Fermi surface.
We employed the quasiclassical approach and
took into account a counterpart to the interlevel spacing of the vortex bound states $\omega_0(\bm{k}_{\rm F})$
as a contribution to the flux-flow resistivity,
in addition to the quasiparticle scattering rate $\varGamma(\bm{k}_{\rm F})$.
The results show that the field-angle dependence of the flux-flow resistivity exhibits different behavior
between the sign conserved line-node $s$-wave pair and the sign reversed $d$-wave one,
irrespective of the fact that those pair potentials have the same amplitude anisotropy.
Therefore, the field-angle dependence of the flux-flow resistivity can be a phase-sensitive probe
if the flux-flow resistivity is observed, for example, by measuring the microwave surface impedance with changing the field-angle direction.

\section*{Acknowledgments}
The authors thank N. Nakai, H. Suematsu,
S. Yasuzuka, Y. Kato, K. Izawa, M. Kato,
A. Maeda, and T. Okada
for helpful discussions.

\end{document}